\documentclass{gHPR2e}

\usepackage{subfigure}
\usepackage{graphicx}
\usepackage{color}

\newcommand{\lbcoe}{La$_{1.875}$Ba$_{0.125}$CuO$_4$}
\newcommand{\lbcoa}{LBCO$_{1/8}$}
\newcommand{\lbco}{La$_{2-x}$Ba$_{x}$CuO$_4$}

\begin{document}


\title{Combined single crystal polarized XAFS and XRD at high pressure: probing the interplay between lattice distortions and electronic order at multiple length scales in high $T_c$ cuprates}

\author{G. Fabbris$^{a,b,c,\ast}$\thanks{$^\ast$Corresponding authors. Emails: gfabbris@bnl.gov, haskel@aps.anl.gov
\vspace{6pt}}, M. H\"ucker$^c$, G. D. Gu$^c$, J. M. Tranquada$^c$, D. Haskel$^{a,\ast}$
\\\vspace{6pt}  $^{a}${\em{Advanced Photon Source, Argonne National Laboratory, Argonne, IL 
60439, USA}}; $^{b}${\em{Department of Physics, Washington University, St. Louis, MO 
63130, USA}}; $^{c}${\em{Condensed Matter Physics and Materials Science Department, Brookhaven National Laboratory, Upton, New York 11973, USA}}}

\maketitle

\begin{abstract}
Some of the most exotic material properties derive from electronic states with short correlation length ($\sim$10-500~\AA), suggesting that the local structural symmetry may play a relevant role in their behavior. Here we discuss the combined use of polarized x-ray absorption fine structure and x-ray diffraction at high pressure as a powerful method to tune and probe structural and electronic orders at multiple length scales. Besides addressing some of the technical challenges associated with such experiments, we illustrate this approach with results obtained in the cuprate \lbcoe, in which the response of electronic order to pressure can only be understood by probing the structure at the relevant length scales.


\end{abstract}

\section{Introduction}

The physical properties of solids are commonly derived from their long-range crystalline structure, as this approach allows for calculating the electronic properties using a small number of atoms contained in the unit cell. However, there is increasing technological and scientific interest in materials for which the assumption of long-range order does not necessarily hold, the most straightforward example being the large scale use of amorphous semiconductors in electronics applications \cite{Kamiya2010,Berthier2011}. Additionally, emergent phenomena such as superconductivity and charge/spin order are often related to the presence of nanoscale inhomogeneity, with theories suggesting that such local disorder may play a critical role for the electronic properties \cite{Dagotto2005,Uberuaga2016}. The study of short range phenomena requires the use of appropriate experimental techniques to capture the structural and electronic motifs at the relevant length scales.

X-ray absorption fine structure (XAFS) is a well-established technique to probe the local structural correlations \cite{Lee1981,Rehr2000,Bunker2010}, and its implementation at third and fourth generation high brilliance synchrotron sources makes it conveniently well suited for the small sample volumes of diamond anvil cells (DAC). However, although pioneer high pressure XAFS experiments were performed in the 1970's and 80's \cite{Shimomura1978,Shimomura1987,Alberding1989}, its widespread use has failed to materialize (for a few examples see Refs. \cite{Shimomura1978, Shimomura1987, Alberding1989, DiCicco1996, Pellicer-Porres1999, San-Miguel2000, Sapelkin2000, Sadoc2000, Pellicer-Porres2000, Comez2001, Pellicer-Porres2002, Vaccari2009a, Hong2009, Baldini2010, Baldini2011, Matsumoto2011, Ishimatsu2012, Chen2013, Pellicer-Porres2013, Fabbris2013, Hong2014, Properzi2015}). One main reason is the severe data distortions introduced by diamond Bragg peaks. Different approaches have recently been suggested to tackle this issue, such as the use of nano-polycrystaline diamond anvils \cite{Nakamoto2007, Baldini2011, Matsumoto2011, Ishimatsu2012, Properzi2015} and polycapillary optics \cite{Chen2013}, as well as through careful data analysis \cite{Hong2009,Hong2014}. Additionally, ``proof of concept" experiments have demonstrated the possibility of using differential \cite{Chu2011} and magnetic \cite{Matsumoto2011} XAFS at high pressure. Despite these advances, the coupling between x-ray linear polarization and XAFS (polarized XAFS) has remained largely untapped, its application confined to a few reports \cite{Pellicer-Porres1999, San-Miguel2000, Pellicer-Porres2000, Pellicer-Porres2013, Fabbris2013}. This is a drawback for the community, as polarized XAFS is an important tool to isolate the behavior of particular bonds in complex local structures \cite{Bunker2010}.

In this manuscript, we will discuss the technical challenges and capabilities of high-pressure polarized XAFS on single crystals using our recent work on \lbcoe\ (\lbcoa) as a reference \cite{Fabbris2013,Fabbris2014}. The \lbco\ family is the first discovered high-$T_c$ superconducting cuprate \cite{Bednorz1986}. It was realized early on that while Ba doping leads to superconductivity, a dramatic suppression of $T_c$ occurs near x = 1/8 \cite{Moodenbaugh1988,Hucker2011}. Such suppression was later found to be related to the emergence of charge and spin stripe orders that are pinned to the lattice by a concomitant structural phase transition to the low-temperature-tetragonal phase (LTT) that breaks the four-fold rotational symmetry of the CuO$_2$ planes \cite{Tranquada1995}. Recent high-pressure diffraction work has challenged this relationship, as charge order (CO) persists at pressures beyond the transition to the high-symmetry high-temperature-tetragonal phase (HTT) \cite{Hucker2010}. However, stripe and superconducting order in \lbcoa\ are intrinsically short-ranged properties with correlation and coherence lengths of $\sim$100-600~\AA\ and $\sim$10-20~\AA, respectively \cite{Hucker2011}. It is therefore critical to examine the relationship between these electronic properties and the local structural symmetry. In fact, our combined single crystal polarized XAFS and x-ray diffraction (XRD) experiments at high pressure indicate that the four-fold local symmetry remains broken at high pressure \cite{Fabbris2013, Fabbris2014}. Such persistent LTT-like local structure has the same correlation length as that of CO regions, suggesting that structural and electronic order remain intertwined at high pressure. Therefore, we demonstrate that the proper understanding of the high pressure properties of \lbcoa\ must include probing the structural and electronic order at multiple length scales.

The remainder of the paper is organized as follows. In section 2, the relevant aspects of the XAFS method will be discussed. Particular attention is given to the ability of polarized XAFS in determining the short range order (section 2.1), the technical challenges of performing polarized XAFS experiments at high pressure (section 2.2), and the experimental details of the \lbcoa\ study (section 2.3). The results obtained are presented and discussed in section 3, followed by final remarks in section 4.

\section{Methods}

\subsection{Polarized XAFS}

Here we will focus on the relevant aspects of polarized XAFS and its interpretation. For additional details on the XAFS technique please refer to the many books and reviews on the subject (for instance, see Refs. \cite{Lee1981,Rehr2000,Bunker2010}). The XAFS signal consists of oscillations in the absorption coefficient observed above an absorption edge. For K-edges and single scattering events the XAFS signal can be approximated by:
\[
\chi (k) = -\sum _i 3 (\vec{e}.\vec{R_i})^2 \frac{N_i S_o^2}{kR_i^2} e^{-2 \sigma _i ^2 k^2} e^{-\frac{2 R_i}{\lambda_k}} \lvert f_{eff} (k) \rvert \sin(2 k R_i + \Phi_k)
\]
with $k = \sqrt{\frac{2m_e}{\hbar} ( E - E_0 ) ^2 }$. $E$ and $E_0$ are the photon and absorption edge energies, respectively. The terms in this equation can be split into two categories: electronic and structural. The electron mean free path ($\lambda_k$), scattering amplitude ($f_{eff} (k)$), and phase shift ($\Phi_k$) are electronic properties that are calculated $ab$ $initio$ (the FEFF8 code was used in the present work \cite{Ankudinov1998}). The structural parameters are the number of degenerate scattering paths (or coordination number) ($N_i$), the path's average length ($R_i$), and the Debye-Waller factor which describes the root-mean-square deviation in path length due to static or dynamical/thermal disorder ($\sigma_i$). These, combined with the absorption edge energy and the amplitude reduction factor ($S_o$), are the adjustable parameters that are typically fitted to the data using the least-squared method (the IFEFFIT/HORAE package was used here \cite{Newville2001, Ravel2005}). The first term of the equation above is particularly important for the present work as it modulates the XAFS amplitude of a scattering path according to its relative orientation to the x-ray polarization, thus the contribution of particular bonds can be enhanced or suppressed by appropriately choosing the sample orientation in a single crystal. As discussed below, this polarization dependence is critical to disentangle individual contributions of nearly degenerate distances. Additionally, XAFS is uniquely sensitive to three body correlations through multiple scattering paths. In particular, the intensity of paths with (nearly) forward electron scattering is enhanced by the so-called ``lensing" effect, thus making XAFS specially sensitive to bond angles near 180$^{\rm o}$. Finally, Fourier transform (FT) is a natural tool to interpret the oscillatory nature of $\chi(k)$. The FT of $\chi(k)$ can thus be interpreted as a ``pseudo" partial radial distribution function, with peak positions not directly positioned at $R_i$ but corrected by $\Phi_k$, and peak shape that reflects the k-dependence of various terms affecting the XAFS amplitude.

\begin{figure}[b]
\begin{center}
\includegraphics[width = 15 cm]{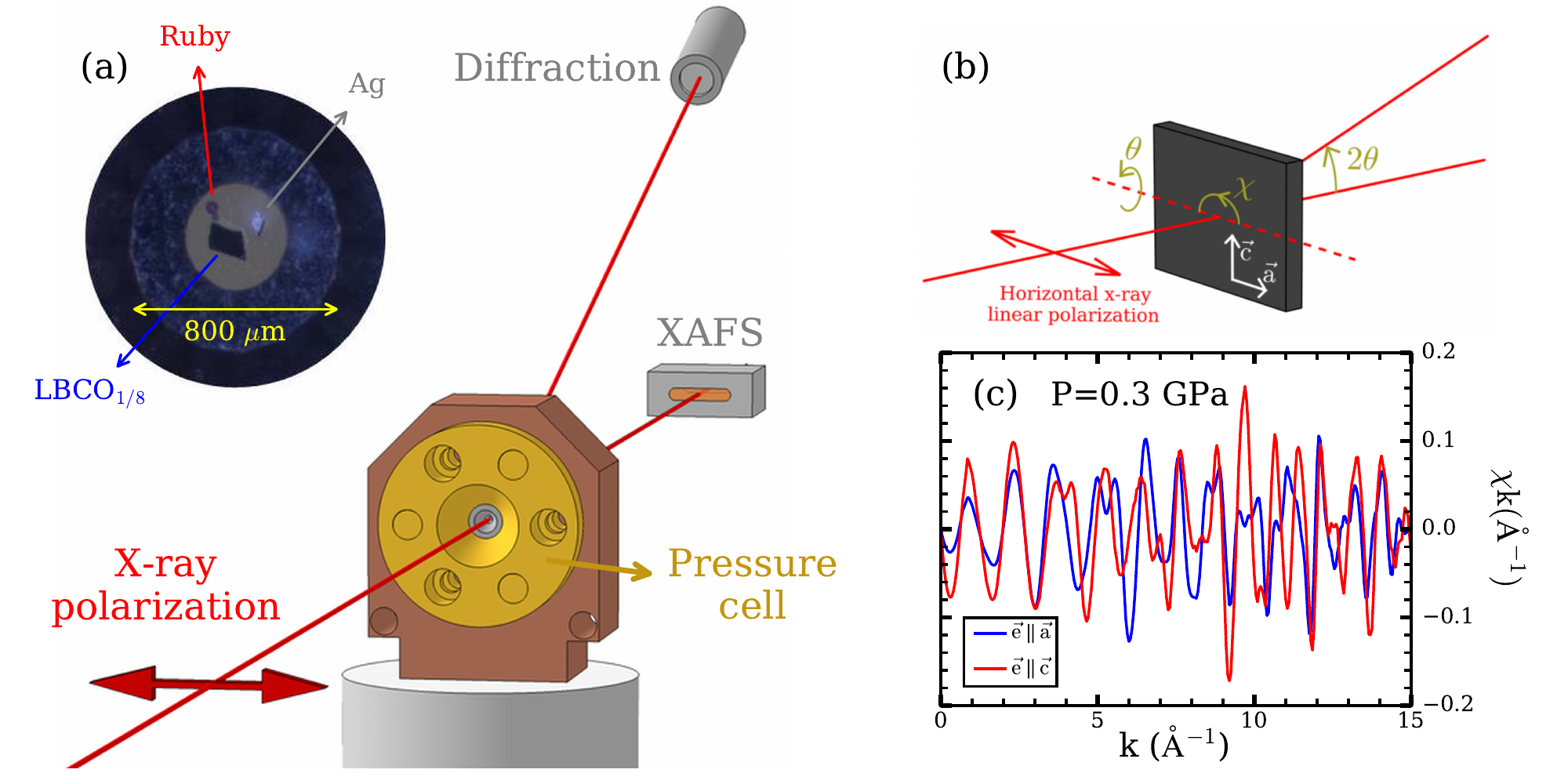} \caption{\label{scheme} (a,b) Experimental setup. The sample is oriented with $ac$ plane normal to the x-ray direction. Therefore, $\vec{a}$ and $\vec{c}$ polarized XAFS is measured by moving $\chi$. (c) $\chi(k)$ measured at 0.3~GPa. The absence of diamond Bragg peaks is critical for accurate XAFS analysis.}
\end{center}
\end{figure}

\subsection{High pressure polarized XAFS}

As with all other high pressure x-ray techniques using single crystalline diamond anvils, polarized XAFS measurements may suffer from the contamination of diamond Bragg peaks. Various effective experimental approaches \cite{Nakamoto2007,Chen2013} as well as data analysis remedies \cite{Hong2009} have been suggested to mitigate this problem, but these are discussed in other sections of this special issue and will not be the focus of this article. The polarized XAFS work displayed here largely benefited from the high sensitivity of the La local environment to phase transitions in \lbcoa\ \cite{Haskel2000} (see section 3). The high energy La K-edge (38.95~keV) strongly reduces the diamond absorption length ($\sim$14~mm, compared to $\sim$1.4~mm around 10 keV), which largely suppresses diamond Bragg peaks intensities. However, even at this energy, a pair of regular diamond anvils ($\sim$4~mm total) still significantly distorted the data. A clean XAFS signal, as seen in Fig. \ref{scheme}(c), could only be obtained by using a partially perforated anvil opposite to a mini-anvil mounted on a fully perforated anvil \cite{Haskel2008}, similar to what was reported in Ref. \cite{Dadashev2001}, which reduces the total diamond thickness to $\sim$0.8~mm.

When compared to other XAFS methods, polarized XAFS adds the experimental constraint of aligning the x-ray linear polarization to the relevant bond directions, which generally requires the use of single crystals. Note that the development of fourth generation synchrotron sources with nanoscale x-ray beams will allow to perform such experiments on a single powder grain. Furthermore, aligned powders have been used in previous ambient pressure work \cite{Haskel1996,Haskel2000}, but it is difficult to conceive using the same approach in a DAC. A few potential issues emerge from using single crystals in XAFS measurements. First, the sample itself may produce Bragg peaks that will distort the spectrum just like those arising from diamond anvils. Although such peaks did not significantly distort our data, it would likely be relevant if lower energy edges are probed. Perhaps the best approach to tackle this problem is to clean the data $a$ $posteriori$ by measuring the XAFS spectrum at different angles as described in Ref. \cite{Hong2009}. Note that even for polarized XAFS, one can select a rotation angle that preserves the alignment of the polarization and crystallographic directions. For instance, in our geometry the alignment of $\vec{e}$ along $\vec{a}$ and $\vec{c}$ is controlled only by $\chi$, thus allowing for the use of $\theta$ in this de-glitching procedure (Fig. \ref{scheme}(b)). Finally, the conventional XAFS measurement is performed in transmission mode, i.e. by measuring the x-ray intensity before and after the sample as shown in Fig. \ref{scheme}(a). However, this setup imposes important constraints on the sample thickness \cite{Lee1981,Bunker2010}. Again, this problem was conveniently addressed here through the use of La K-edge XAFS for which the optimal crystal thickness lies around $\sim$60~$\mu$m. However, optimal transmission thicknesses for most transition-metal oxides are much smaller ($\sim$5-15~$\mu$m), leading to the associated challenges of producing and manipulating such thin single crystals. This issue can be addressed by measuring XAFS in fluorescence mode \cite{Bunker2010}, $e.g.$ with the detector placed in a nearly backscattering geometry.

\subsection{Experimental setup}

\begin{figure}
\begin{center}
\includegraphics[width = 15 cm]{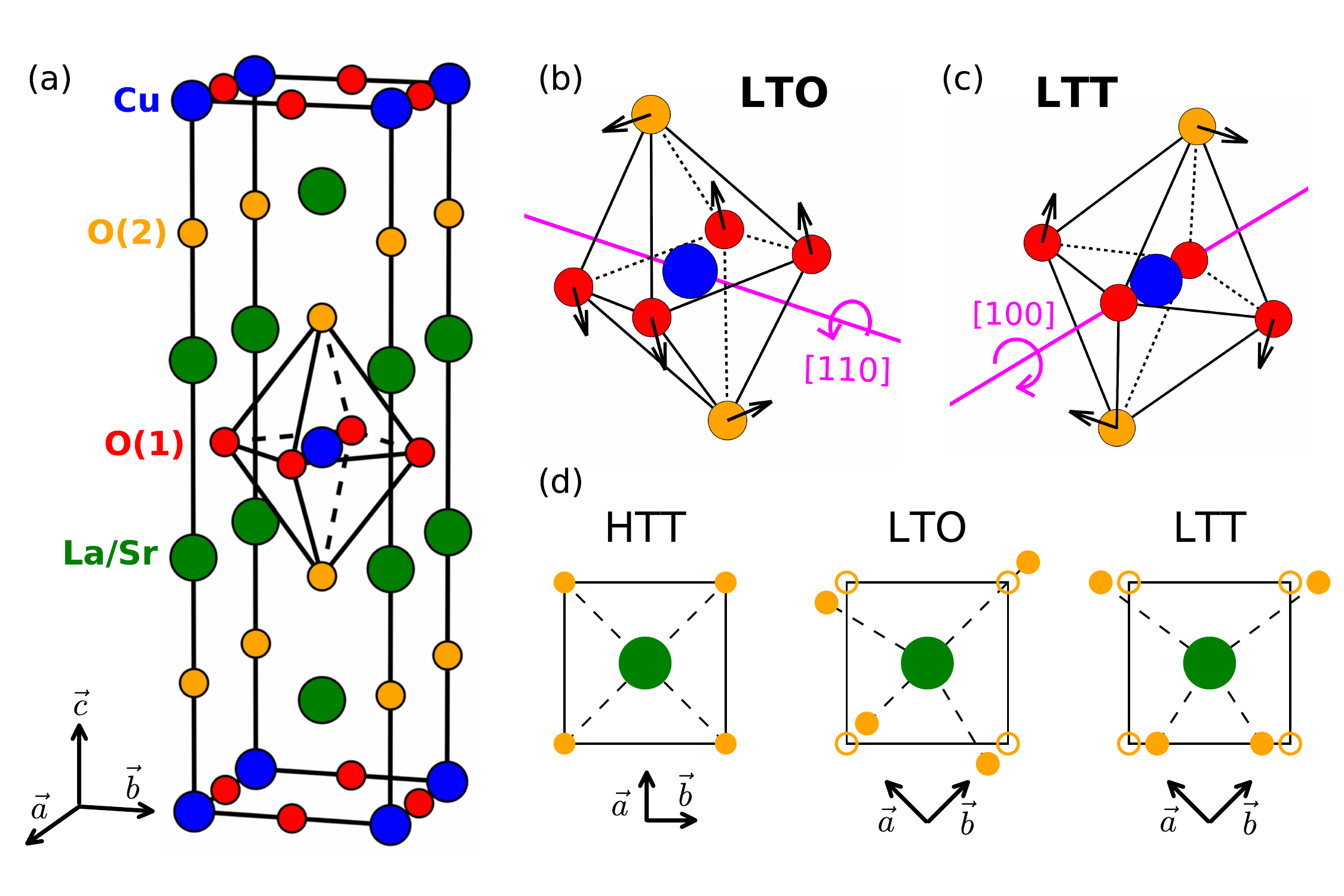} \caption{\label{struct} (a) \lbcoa\ HTT crystal structure. (b,c) The LTO and LTT phases are stable at low temperature (235~K and 54~K, respectively). These correspond to two different rotations of the CuO$_6$ octahedra, the LTO/LTT tilt is along the [110]/[100] direction. (d) The CuO$_6$ tilt strongly distort the bonds in the LaO plane. The four degenerate in-plane La-O(2) bonds of the HTT struture split into three and two different bonds in the LTO and LTT phases, respectively.}
\end{center}
\end{figure}

\lbcoe\ single crystals were grown using the traveling-solvent floating-zone technique. Throughout this manuscript, diffraction Bragg peaks and crystal directions will be labelled according to the HTT unit cell (a = 3.78~\AA\ and c = 13.2~\AA) (Fig. \ref{struct}(a)). Measurements were performed at beamline 4-ID-D of the Advanced Photon Source, Argonne National Laboratory. Horizontally polarized x-rays were produced by a planar undulator. A schematic of the experimental setup is displayed in Fig. \ref{scheme}. XAFS measurements were performed in transmission mode at the La K-edge (38.95 keV) with Ar/Kr filled ion chambers as incident/transmitted intensity detectors. The beam size was set to 50x50~$\mu$m$^2$ using slits, while temperature was kept at 5~K using a closed-cycle cryostat. Good quality data was obtained to $\sim$860~eV ($k \sim$ 15~\AA$^{-1}$) (Fig. \ref{scheme}(c)), leading to a spatial resolution of $\sim$0.1~\AA\ in the refined distances. The (006) and (200) Bragg peaks were measured to extract the lattice parameters and to align the x-ray polarization vector along $\vec{a}$ and $\vec{c}$, respectively. Additionally, the LTT-only (100) peak was measured to verify the previously reported pressure-induced LTT-HTT phase transition \cite{Hucker2010}. The pressure cell was assembled with a partially perforated anvil opposite to a mini-anvil as reported in Ref. \cite{Haskel2008}, and similar to the configuration in Ref. \cite{Dadashev2001}. Anvils with 0.6~mm culet diameter were used with stainless-steel gaskets and 4:1 methanol:ethanol pressure media. Pressure was applied at low temperature using a He-gas membrane. A small ruby ball was used to check the pressure right after loading, and silver diffraction was used as the in-situ manometer \cite{Holzapfel2001}. 

Additional x-ray diffraction measurements were performed to investigate the evolution of CO and LTT domains at high pressure. This experiment was performed with the same setup described above, but 20 keV photons were used in order to take advantage of the beamline's optimal x-ray flux. CO and local LTT domains were studied by analyzing the (2-2$\delta$ 0 1/2) and (3/2 3/2 2) Bragg peaks ($\delta \sim 0.125$), respectively.

\section{Results and discussion}

Three crystal structures are observed in \lbcoa\ as a function of temperature (Fig. \ref{struct}) \cite{Axe1989,Hucker2011}. The high-temperature tetragonal (HTT) phase is stable above 235~K. This phase features unrotated CuO$_6$ octahedra, leading to a four-fold symmetry of the CuO$_2$ plane. Upon cooling below 235~K, the CuO$_6$ octahedra rotates about the [110] direction (Fig. \ref{struct}(b)), driving the low temperature orthorombic (LTO) phase. At 54~K the octahedral rotation axis switches to the [100] direction (Fig. \ref{struct}(c)), leading to the low-temperature tetragonal (LTT) structure. CO emerges concomitantly to the LTO-LTT phase transition. As discussed in section 2.2, the high energy of the La K-edge is particularly favorable for XAFS measurements in a DAC. Additionally, LTO and LTT CuO$_6$ rotations drive a dramatic displacement of apical oxygens (O(2)) within the $ab$ plane, which in turn strongly distorts ($> 0.15 \rm \AA$) the once degenerate in-plane La-O(2) bonds (Fig. \ref{struct}(d)). However, La's first coordination shell consists of a total of nine La-O bonds. Therefore, the XAFS ability to detect the La-O(2) bond distortion driven by CuO$_6$ rotation largely relies on isolating the $a$ and $c$ axis contributions using linearly polarized x-rays \cite{Haskel2000}. Placing the x-ray linear polarization along $\vec{a}$ selects in-plane La-O(2) and out of plane La-O(1) bonds (La-O(1) is $\sim$45$^{\rm o}$ away from the $ab$ plane) (Fig. \ref{struct}(a)). Alternatively, when $\vec{e} \parallel \vec{c}$, out of plane La-O(1) and La-O(2) bonds are probed. The in-plane La-O(2) behavior is thus isolated by concomitantly fitting both measurements.

\begin{figure}
\begin{center}
\includegraphics[width = 15 cm]{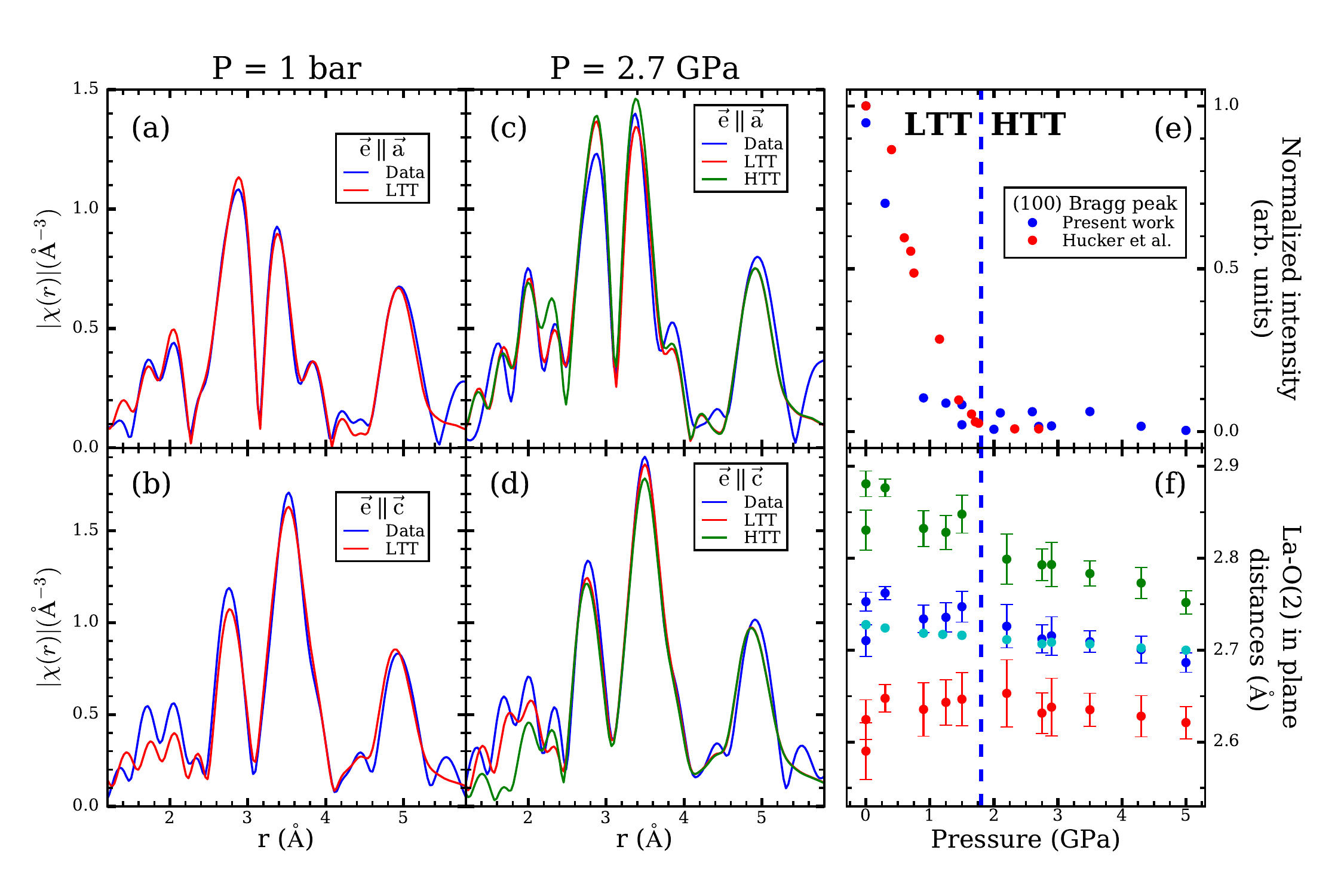} \caption{\label{exafs} (a,b) Ambient pressure XAFS experimental data and LTT model for $\vec{e}$ along $\vec{a}$ and $\vec{c}$. (c,d) XAFS experimental data beyond LTT-HTT transition (2.7~GPa) compared to LTT and HTT models for $\vec{e}$ along $\vec{a}$ and $\vec{c}$. (e) Intensity of the LTT-only (100) peak demonstrating the emergence of long range HTT symmetry, in agreement with previous results \cite{Hucker2010}. (f) La-O(2) distances measured by XAFS (green and red). In blue and cyan are the average La-O(2) distance measured by XAFS and XRD, respectively.}
\end{center}
\end{figure}

Figure \ref{exafs} summarizes our main findings. At ambient pressure the long range LTT symmetry is also captured in the local structure (Fig. \ref{exafs}(a,b)), in agreement with previous XAFS and PDF work \cite{Billinge1994,Haskel2000,Bozin2015}. The pressure dependence of the (100) diffraction peak also reproduces previous work (Fig. \ref{exafs}(e)) \cite{Hucker2010}, demonstrating that the XAFS measurements were indeed performed across the macroscopic (long range) LTT-HTT phase transition. As discussed earlier, the in-plane La-O(2) bonds are the most sensitive to CuO$_6$ tilts, contributing only to the $\vec{e} \parallel \vec{a}$ XAFS signal. As seen in Fig. \ref{exafs}(c,d), the LTT model is far superior to the HTT model in describing the La-O bonds signal even within the macroscopic HTT phase, demonstrating that the local symmetry remains LTT even when the long range order is HTT. This conclusion is supported by the significantly smaller R-factor obtained using the LTT (4.3\%) versus the HTT (13.6\%) model. Additionally, an HTT fit yields an unphysical increase in Debye-Waller factor ($\sigma ^{2.7GPa}_{HTT} \approx 0.3 \rm \AA$, compared to $\sigma ^{1bar} \approx 0.05 \rm \AA$). A small reduction in $\sigma$ is observed for the LTT model ($\sigma ^{2.7GPa}_{LTT} \approx 0.03 \rm \AA$) as expected in a contracted lattice. Finally, while LTT tilts give rise to a particular distribution of La-O(2) distances, the average La-O(2) bond distance is approximately the same as in the HTT phase. In order to further verify our results, the pressure dependence of the average La-O(2) distance was calculated using both XAFS and XRD (lattice parameters were extracted from the (006) and (200) Bragg peaks) (Fig. \ref{exafs}(f)). The close agreement of the average in-plane La-O(2) distance measured by long and short range probes provides strong support to our conclusion that the high-pressure HTT phase of \lbcoa\ is composed of local LTT symmetry.

\begin{figure}
\begin{center}
\includegraphics[width = 15 cm]{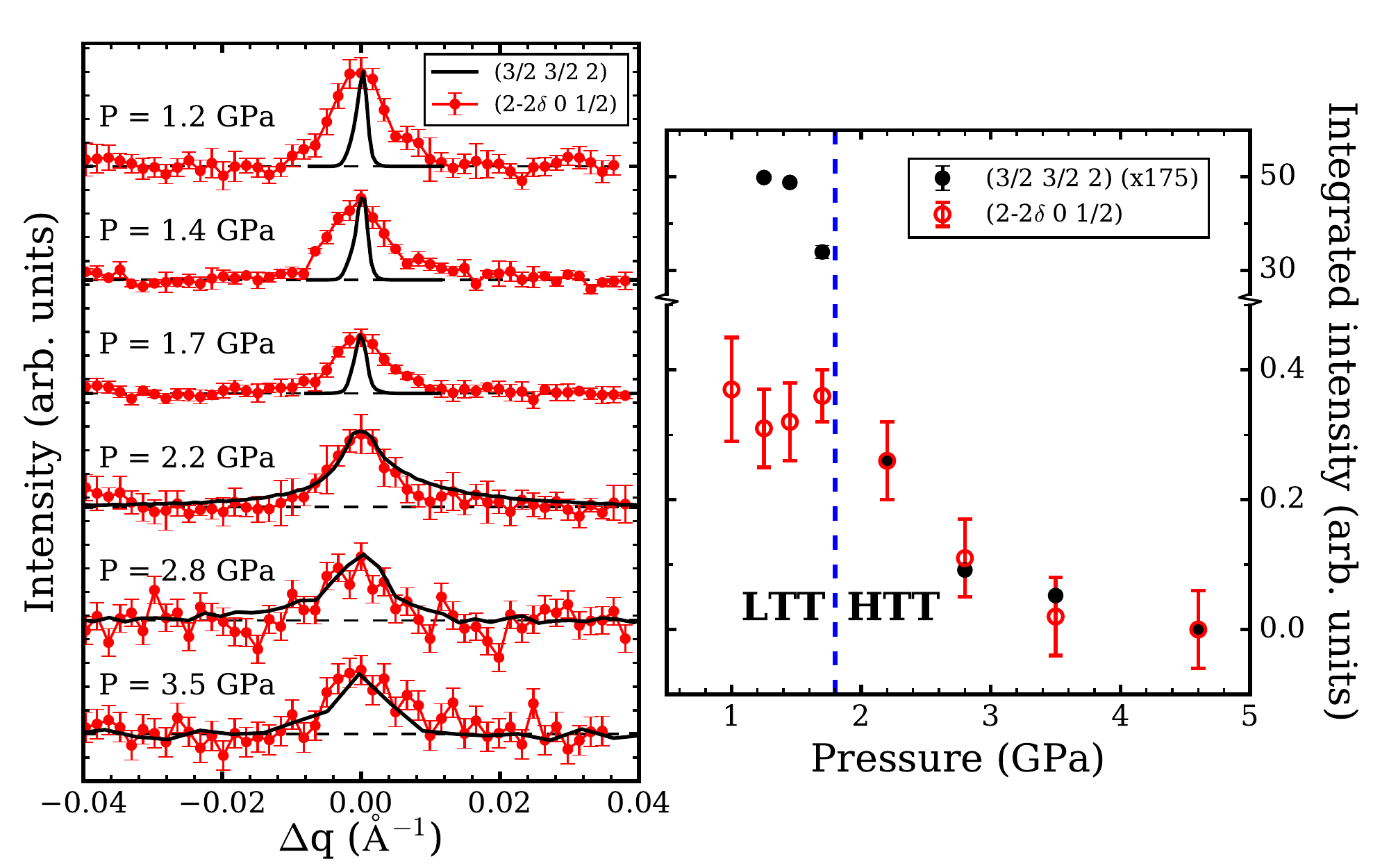} \caption{\label{xrd} The pressure dependence of (3/2 3/2 2) and CO (2-2$\delta$ 0 1/2) Bragg peaks is displayed on the left panel. Intensities are re-scaled for clarity. The correlation length of (3/2 3/2 2) dramatically changes across the LTT-HTT transition. The right panel shows the integrated intensity of these peaks as a function of pressure. While (3/2 3/2 2) peak intensity is strongly suppressed across the LTT-HTT transion, no clear discontinuity is observed on the CO Bragg peak intensity. Pressures beyond 2.2~GPa concomitantly suppress both Bragg peaks.}
\end{center}
\end{figure}

Although polarized XAFS results demonstrate the persistence of local LTT tilts at high pressure, it contains no information on how such LTT tilts correlate across the longer length scale spanned by CO regions ($\sim$200~\AA). Following previous studies \cite{Hucker2010}, the high pressure behavior of the (3/2 3/2 2) and CO (2-2$\delta$ 0 1/2) Bragg peaks was investigated ($\delta \sim$ x = 0.125). The CO peak displays a nearly constant intensity up to 2.2~GPa (above LTT-HTT transition), being gradually suppressed at higher pressures and becoming undetectable beyond 3.5~GPa (Fig. \ref{xrd}). Noticeably, the CO correlation length remains nearly constant in the whole range. The strong reduction in intensity of the (3/2 3/2 2) peak is commensurate with the suppression of long range LTT order. However, while the width of (3/2 3/2 2) in the LTT phase is determined by the instrumental resolution, it severely broadens at the onset of the high-pressure HTT phase (Fig. \ref{xrd}). Remarkably, not only do the high-pressure LTT domains display the same correlation length of the CO domains, but phase correlations are also concomitantly suppressed above 3.5~GPa.

\begin{figure}
\begin{center}
\includegraphics[width = 15 cm]{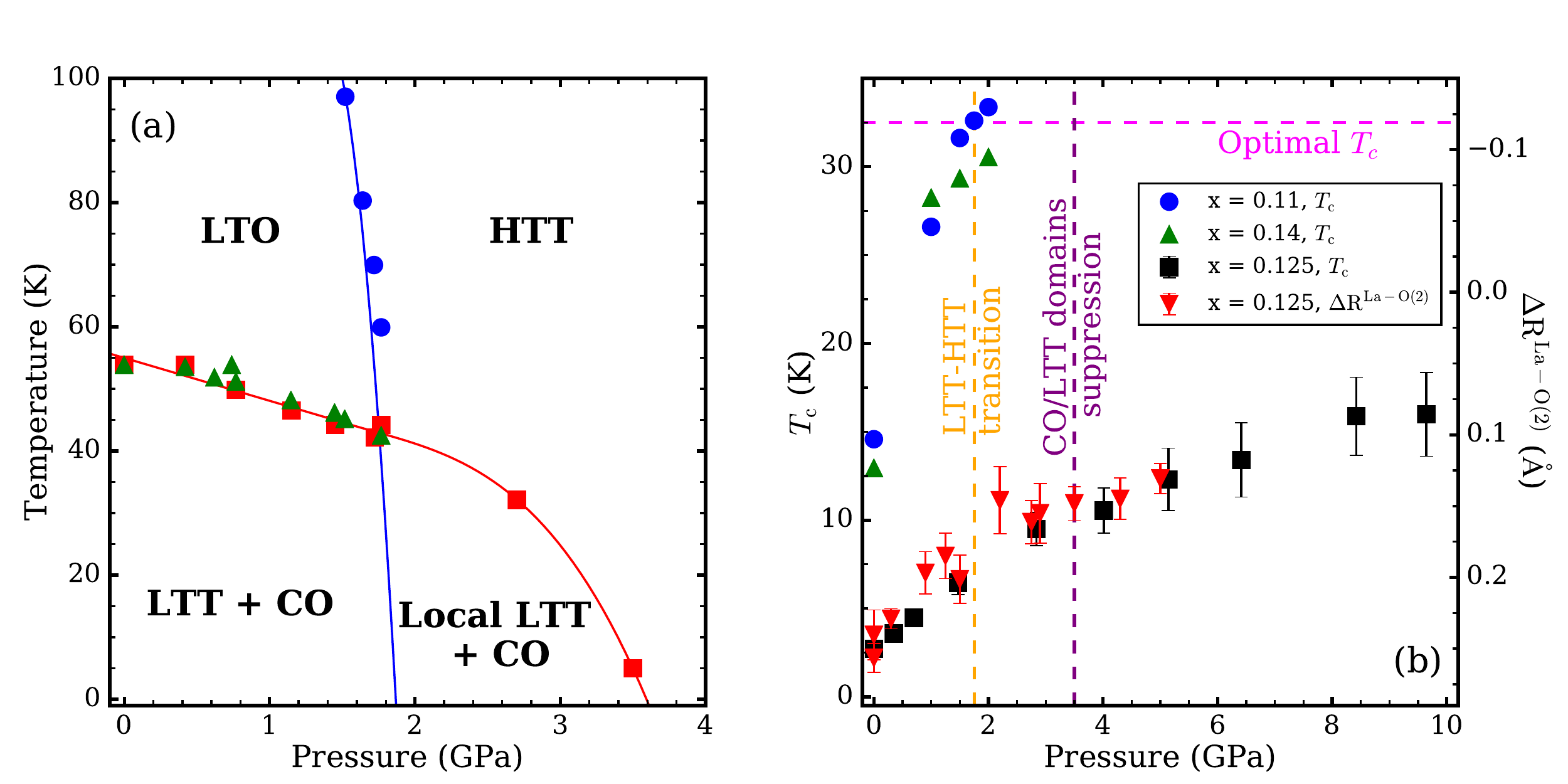} \caption{\label{diagrams} (a) Revised \lbcoa\ phase diagram which includes the electronic and structural behavior of local-, medium- and long- range order. Includes data from Ref. \cite{Hucker2010}. (b) Pressure dependence of the superconducting $T_c$ of \lbco\ for x = 0.11, 0.125, 0.14 \cite{Yamada1992,Hucker2010} is plotted together with $\Delta$R$^{\rm La-O(2)}$ measured by XAFS.}
\end{center}
\end{figure}

A revised phase diagram for \lbcoa\ is therefore needed to accommodate the XAFS and XRD results (Fig. \ref{diagrams}(a)). The onset of the long-range HTT order only weakly couples to the local LTT tilts and CO domains. Indeed, such high-pressure HTT phase is actually composed of local LTT domains whose correlation length matches that of CO. Combined with the concomitant suppression of CO and LTT domains, this result strongly supports a very close connection between structural and electronic local order in \lbcoa. However, the origin of the high pressure LTT domains remains unclear. They may simply persist due to defects or non-hydrostaticity, but it can also be driven by the strong electron-lattice coupling, which may force the presence of LTT domains in order to retain CO. Furthermore, the local structure probed by XAFS remains LTT-like even upon the suppression of LTT domains. XAFS is an ``instantaneous" probe ($\sim$1~fs time scale), therefore it is unclear if the persisting short range LTT correlation is static or dynamic above 3.5~GPa. A complete mapping of the CO and LTT domains phase diagram is needed to address both their origin and relationship with the pressure enhanced superconducting $T_c$; see Fig. \ref{diagrams}(b).

Finally, we address the anomalous pressure dependence of $T_c$ seen in \lbco\ at x = 1/8 (Fig. \ref{diagrams}(b)). Although suppressed at ambient pressure, $T_c$ reaches values comparable to optimally doped samples within $\sim$2~GPa for Ba doping slightly away from 1/8 (x = 0.11 and 0.14) \cite{Yamada1992}. However, for x = 1/8 even 15~GPa drives $T_c$ to only $\sim$18~K \cite{Hucker2010}. Such disparate pressure dependence is very puzzling, as purely electronic effects such as charge transfer and band broadening are expected to be independent of doping. Additionally, the pressure dependence of  $T_c$ in \lbcoa\ appears to be insensitive to the LTT-HTT transition or to the suppression of LTT and CO domains. Remarkably, $T_c$ scales with the reduction of the local LTT tilt amplitude observed at high pressure (Fig. \ref{diagrams}(b)). The suppression of superconductivity in \lbcoa\ is believed to occur due to a frustration of the Josephson coupling along $\vec{c}$ driven by the 90$^{\rm o}$ rotation of LTT tilts between neighboring CuO$_2$ planes \cite{Himeda2002,Berg2009}. Therefore, the presence of local LTT tilts might be sufficient to stabilize stripe order, which in turn could suppress superconductivity. This picture is supported by the very short superconducting coherence length ($\sim$10-20~\AA) seen in cuprates. A combined high pressure XAFS and XRD study of dopings away from 1/8 is needed to clarify this scenario.

\section{Final remarks}

In this manuscript we discussed technical challenges and methodologies for the use of polarized XAFS at high pressure using single crystals in diamond anvil cells. We show that combining polarized XAFS and XRD at high pressure allows probing the structural and electronic properties at multiple length scales, a critical capability to fully comprehend the role of local disorder in materials properties. 

We demonstrate the potential of this methodology through our work on the cuprate \lbcoa. A combination of polarized XAFS and XRD is used to show that the high-pressure high-symmetry HTT phase is actually composed of local low-symmetry LTT order. In the intermediate length scale, this high-pressure local LTT symmetry forms domains of $\sim$200~\AA, which are tightly correlated to the behavior of CO domains, i.e. they display the same correlation length and are concomitantly suppressed above 3.5~GPa. Additionally, the slow increase of $T_c$ at high pressure appears to correlate with the slow reduction of LTT distortion observed with polarized XAFS, suggesting the presence of local CO correlations even at pressures in which CO domains could not be detected. These results strongly suggest that the electronic properties of high-$T_c$ cuprates should be examined in light of the structural motif at the relevant length scales. 

The future of high-pressure polarized XAFS is promising. The results displayed here are a testament to the ability of high-pressure polarized XAFS to help understand emergent material properties at high pressure. Recent technical developments are key in enabling the use of polarized XAFS in a broad set of materials. We particularly point out the use of nano-polycrystalline diamond anvils, which, despite being used for almost 10 years \cite{Nakamoto2007}, have only recently become accessible to a wider community. Additionally, the emergence of fourth generation synchrotron sources delivering sub-micrometer x-ray beams will open up the possibility of using polarized XAFS even on a single powder grain, removing most of the technical challenges associated with producing high-quality single crystals. These developments will likely place polarized XAFS on a par with other more established tools in high-pressure research.

\section*{Acknowledgments}
Work at Argonne National Laboratory (Brookhaven National Laboratory) was supported by the U.S. Department of Energy, Office of Science, Office of Basic Energy Sciences, under Contract No. DE-AC02-06CH11357 (DE-SC00112704). The Advanced Photon Source at Argonne National Laboratory is a U.S. Department of Energy Office of Science User Facility. G.F. was also supported by the U.S. Department of Energy under Award Number 1047478. We would like to thank Yejun Feng for his valuable advice and help during the experiment.

\bibliographystyle{gHPR}
\bibliography{lbco_ref}

\end{document}